\documentclass[acmsmall]{acmart}
\AtBeginDocument{%
  \providecommand\BibTeX{{%
    \normalfont B\kern-0.5em{\scshape i\kern-0.25em b}\kern-0.8em\TeX}}}

\usepackage{microtype}
\usepackage{graphicx}

\usepackage{color,soul}
\usepackage{caption,subfigure}
\usepackage{latexsym,tabularx}
\usepackage{enumitem}
\setlist{leftmargin=3mm}
\usepackage{amsfonts} % Math packages
\usepackage{tikz,amsmath,amsthm,graphicx}
\usepackage{varwidth,booktabs}
\usepackage{xcolor}
\usepackage{algorithm,algpseudocode}
\usepackage{array, makecell}
\usepackage{multirow}
\usepackage{tabularx}
\usepackage{mwe}
\usepackage{capt-of}

\usepackage{hyperref}
\usepackage{tikz}
\usepackage{verbatim}
\usepackage{smartdiagram}
\usepackage{booktabs}
\usepackage{tabularx}

\begin{document}

%%\title{Combating Disinformation in Computational Medicine: Characterization, Detection, and Intervention}
% \title{Combating Misinformation in Computational Healthcare: Characterization, Detection, and Intervention}

\title{Combating Healthcare Misinformation in Social Media: Characterization, Detection, and Intervention}

\title{Combating Health Misinformation in Social Media: Characterization, Detection, and Intervention}

\title{Combating Health Misinformation in Social Media: Characterization, Detection, Intervention, and Open Issues}

% \title{Combating Health Misinformation in Social Media: Characterization, Detection, Intervention and Open Issues}


\author{Canyu Chen}
\authornote{Equal contributions.}
\email{cchen151@hawk.iit.edu}
\affiliation{%
  \institution{Department of Computer Science, Illinois Institute of Technology}
  \city{Chicago}
  \state{IL}
  \country{USA}
}

\author{Haoran Wang}
\authornotemark[1]
\email{hwang219@hawk.iit.edu}
\affiliation{%
  \institution{Department of Computer Science, Illinois Institute of Technology}
  \city{Chicago}
  \state{IL}
  \country{USA}
}

\author{Matthew Shapiro}
\email{shapiro@iit.edu}
\affiliation{%
  \institution{Department of Social Science, Illinois Institute of Technology}
  \city{Chicago}
  \state{IL}
  \country{USA}
}

\author{Yunyu Xiao}
\email{yux4008@med.cornell.edu}
\affiliation{%
  \institution{Department of Population Health Sciences, Weill Cornell Medicine}
  \city{New York}
  \state{NY}
  \country{USA}
}

\author{Fei Wang}
\email{few2001@med.cornell.edu}
\affiliation{%
  \institution{Department of Population Health Sciences, Weill Cornell Medicine}
  \city{New York}
  \state{NY}
  \country{USA}
}

\author{Kai Shu}
\email{kshu@iit.edu}
\affiliation{%
  \institution{Department of Computer Science, Illinois Institute of Technology}
  \city{Chicago}
  \state{IL}
  \country{USA}
}

%%\renewcommand{\shortauthors}{Trovato and Tobin, et al.}

\begin{abstract}
Social media has been one of the main information consumption sources  for the public, allowing people to seek and spread information more quickly and easily. However, the rise of various social media platforms also enables the proliferation of online misinformation. In particular, misinformation in the health domain has significant impacts on our society such as the COVID-19 infodemic. Therefore, health misinformation in social media has become an emerging research direction that attracts increasing attention from researchers of different disciplines. Compared to misinformation in other domains, the key differences of health misinformation include the potential of causing actual harm to humans' bodies and even lives, the hardness to identify for normal people, and the deep connection with medical science. In addition, health misinformation on social media has distinct characteristics from conventional channels such as television on multiple dimensions including the generation, dissemination, and consumption paradigms. Because of the uniqueness and importance of combating health misinformation in social media, we conduct this survey to further facilitate interdisciplinary research on this problem. In this survey, we present a comprehensive review of existing research about online health misinformation in different disciplines. Furthermore, we also systematically organize the related literature from three perspectives: characterization, detection, and intervention. Lastly, we conduct a deep discussion on the pressing open issues of combating health misinformation in social media and provide future directions for multidisciplinary researchers.
\end{abstract}



\keywords{misinformation, health, social media}

\maketitle

% \subsection{The Concept of Healthcare Misinformation}
% \cy{TODO: add a figure for the relation between misinformation \& other concepts}

% \cy{explain concepts}

% \cy{ challenge => WHY from these three perspectives => open issues}

% \input{sec_introduction.tex}
% \input{sec_characterization_new.tex}
% \input{sec_detection_new.tex}
% \input{sec_intervention_new.tex}
% \input{sec_dataset.tex}
% \input{sec_future}
% \input{sec_conlusion.tex}

%%%%%%%%%%%%%%%%%%%%%%%%%%%%%%%%%%%%%%%%%%%%%%%%%%%%%%%%%%%%%%%%%%%%%%%%%%%%%%%%%%%%%%%%%%%%%%%%%%%%%%%%%%%%%%%%%%%%%%%%%%%

\section{Introduction}\label{sec:intro}

% \cite{office2021confronting}

% misinformation in social media is a big problem
Social media has become the leading platform for individuals to communicate~\cite{perrin2015social}. People are used to seeking information online and social media platforms can spread information quickly and at scale~\cite{al2019viral,shearer2020news}. However, it can also facilitate the rampant propagation of misinformation and disinformation~\cite{shu2019detecting,allcott2017social}, which has dramatically affected many aspects of our lives. In particular, in the health domain, misinformation has significant impacts on human health and economic loss. 
% health misinformation is under studied and a deep investigation is needed
% Although health misinformation is not new, today’s environment is different. 
For example, the COVID-19 infodemic has been an unprecedented challenge because we are experiencing an epidemic in a digitized and globalized society~\cite{calleja2021public}. Other types of health misinformation are also disseminated widely such as vaccine advertising and propaganda~\cite{broniatowski2018weaponized, jamison2020vaccine}, misleading posts about Zika virus~\cite{sharma2017zika, ghenai2017catching}, false claims on tobacco, vaping, and marijuana products~\cite{albarracin2018misleading}, conspiracy theories on 5G-Corona~\cite{10.1145/3472720.3483617,pogorelov2020fakenews}, which pose a serious threat to the public. To help mitigate the negative effects caused by health misinformation–-both to benefit the public and the health ecosystem, there is a pressing need to combat health misinformation in social media.

Compared to misinformation in other domains such as politics and entertainment, health misinformation has multiple key differences. \textit{First}, health misinformation is directly connected with humans' physical health and can cause actual harm to the body and even lives. Typically, health misinformation is shown to mislead the public to use bleach as COVID-19 prophylaxis and holds a deep association with a spike in bleach poisoning\cite{chary2021geospatial}. \textit{Second},  normal people are less likely to have the ability to identify health misinformation, as it usually requires specialized medical knowledge and expertise. For example, if a piece of misinformation consists of many medical terminologies, only health experts with specific domain knowledge have the capacity  to distinguish it from real information. \textit{Third}, different from  misinformation from other domains such as politics, which may be manipulated narratives and controversial among different groups of people~\cite{shu2017fake}, most health misinformation can be supported by time-tested medical science. Thus, the debunking and prevention of health misinformation can be more convincing to the general public.

Furthermore, health misinformation on social media holds some unique characteristics.
Although health misinformation is not new--health misinformation has been disseminated via conventional channels such as television and radio for years~\cite{lowinger1977health,gollust2019television}, the rise of a variety of social media platforms has completely transformed the \textit{generation}, \textit{dissemination} and \textit{consumption} paradigms of health misinformation, especially during the time of public health crisis such as COVID-19 pandemic~\cite{firoj_covid:icwsm21}. For the \textit{generation} aspect, different from other conventional dissemination channels where misinformation is originated from deliberate producers, a large amount of health misinformation on social media is generated by users, which includes various types such as posts, messages, images and videos~\cite{darwish2022survey}.  For the \textit{dissemination} aspect, as for health misinformation, the transmission speed  is faster and the impact scope is broader due to the inherent networking structure of social media  platforms~\cite{safarnejad2020contrasting}.  For the \textit{consumption} aspect, social media users are more likely to accept online health misinformation owing to the echo chamber and herd behavior effects~\cite{cinelli2021echo,tang2014mining}.

In this article, we present an overview of the research on combating health misinformation in social media and discuss open issues and promising future directions. The motivations of this survey can be summarized as follows: (1) Since online health information has become an important source for the public to acquire medical advice~\cite{vyas2021proliferation}, it is essential to study the impact of various types of health misinformation on humans and society. This review can provide an overview of the dramatic impact and potential harms of online health misinformation.  (2) The unique characteristics of health misinformation in social media have brought new challenges for detection and intervention. A review is urgently desired to systematically analyze the existing efforts in combating online health misinformation. (3) Considering the research on health misinformation in social media involves works from different communities and backgrounds, especially bioinformatics and public health,  which is distinct from misinformation research in other domains, a review is strongly needed to include efforts from different views and standpoints to promote interdisciplinary communication, understanding, and collaboration.  (4) The research on online health misinformation is still in the early stage, and there are many emerging challenges that need further investigation. It is necessary to discuss potential research directions on combating health misinformation in social media.

To this end, to facilitate the research of combating health misinformation on social media in multidisciplinary communities including sociology, psychology, bioinformatics, public health, social computing, data mining and artificial intelligence, we  review the existing research on online health misinformation in different research communities from three perspectives: \textit{characterization}, \textit{detection} and \textit{intervention}. Lastly, we discuss the pressing \textit{open issues} to be addressed through interdisciplinary efforts in the future. Our key contributions to this survey are as follows: (1) We conduct a comprehensive literature review and  cover the viewpoints from multiple disciplines on combating health misinformation in social media. (2) We organize the existing research about online health misinformation from  characterization, detection, and intervention perspectives in a principled and systematical way.  (3) We conduct a deep discussion on multiple open issues of combating online health misinformation and provide future directions for multidisciplinary research communities.

%%%%%%%%%%%%%%%%%%%%%%%%%%%%%%%%%%%%%%%%%%%%%%%%%%%%%%%%%%%%%%%%%%%%%%%%%%%%%%%%%%%%%%%%%%%%%%%%%%%%%%%%%%%%%%%%%%%%%%%%%%%

\section{Health Misinformation Characterization} \label{char}

% \noindent In today's digital age, our society not only needs to deal with public health crises such as the COVID-19 pandemic, but also the increasing misinformation associated with it. To some level, health misinformation could spread just as fast as the virus itself and pose a significant threat to the public. The world health organization (WHO) used the word ``infodemic" \cite{world2021infodemic} to describe the explosion of false or misleading information during a disease outbreak as well as the negative impacts that come along with the spread of misinformation. 

In this section, we address the causes of health misinformation as well as its detrimental effects. We discuss the unique challenges of dealing with health misinformation. Specifically, we first characterize the causes and consequences of health misinformation from different perspectives. Then we highlight the influence strategies and the dissemination patterns of health misinformation.

% \cy{Characterization method: societal methods and data-driven methods}
% \cy{Four types of misinformation appear in online health communities: advertising, propaganda, misleading information, and unrelated information.
% \hw{I feel like this article is more suited for section 3, the article talked about incorporating behavioral features into machine learning. The four types of misinformation are for datasets, they didn't mention the characteristics of each type. I have covered advertising, propaganda, and misleading information in this section.}
% \cite{zhao2021detecting}}

%% Credit: https://texample.net/media/tikz/examples/TEX/work-breakdown-structure.tex
\usetikzlibrary{mindmap,trees}
\begin{figure}[!htbp]
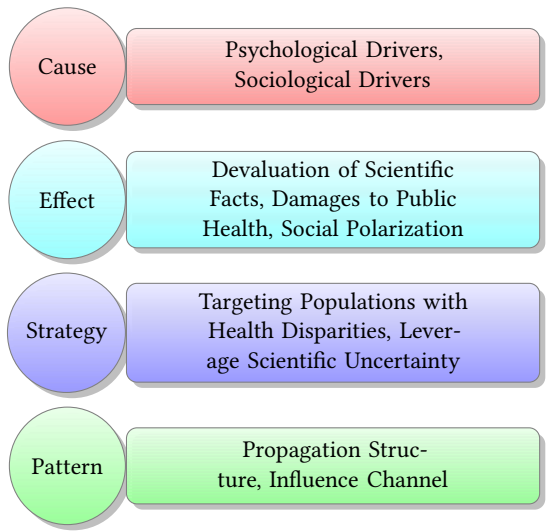

    \centering
       \smartdiagram[descriptive diagram]{
          {Cause,{Psychological Drivers, Sociological Drivers}},
          {Effect, {Devaluation of Scientific Facts, Damages to Public Health, Social Polarization}},
          {Strategy, {Targeting Populations with Health Disparities, Leverage Scientific Uncertainty}},
          {Pattern, {Propagation Structure, Influence Channel}}}
    \caption{Concept map of Health Misinformation Characterization}
    \label{characterization}
\end{figure}

\subsection{Causes of Health Misinformation}
In order to provide effective detection and intervention strategies, we need first to identify the root causes of the spread of health misinformation. We describe the two main drivers for the propagation of health misinformation from psychological and sociological standpoints.

\noindent \textbf{Psychological Drivers of Health misinformation.}
Though health misinformation is not new, the COVID-19 pandemic has brought it to the public's attention to a large degree. Despite the clear divergence from commonsense, false narratives such as ``COVID-19 was caused by 5G technology'' were spread widely. This phenomenon is caused by psychological and cognitive reasons. Uscinski et al.~\cite{uscinski2020people} investigated the psychological foundations of COVID-19 conspiracy beliefs, which consist of two psychological predispositions, denialism and conspiracy thinking. Their experiments showed that these psychological factors make it difficult to correct conspiracy beliefs among groups that exhibit high levels of conspiracy thinking and denialism. 

In addition, Liden et al.~\cite{van2022misinformation} concluded that there are two psychological and cognitive reasons for people being susceptible to misinformation. The first one is the illusory truth effect \cite{southwell2018misinformation}, which refers to the occurrence that repeated claims are more likely to be judged as true than non-repeated claims. This is caused by the cognitive mechanism called processing fluency \cite{wang2016known}. Specifically, processing fluency stated that the more a claim is repeated, the more familiar it becomes and the easier it is to process. The second reason is the inattention reasoning behavior people encounter when processing information. Human cognition theory suggests that people rely on two different processes of reasoning. The first process is mostly based on intuition and is often erroneous, and the second process is based on reflection. Therefore, it is crucial for people to not solely rely on their intuition. Cognitive reflection \cite{frederick2005cognitive} refers to the ability to suppress intuition in favor of reflection during the thinking process. In a separate study, Ali and Qazi \cite{ali2022cognitive} found a positive correlation between higher cognitive reflection and greater truth discernment.

Lastly, Scherer et al. \cite{scherer2021susceptible} tested four psychological hypotheses that attribute to people who are susceptible to online health misinformation. These hypotheses are ``deficits in knowledge or skill, preexisting attitudes, trust in health care and science, and cognitive miserliness.'' Their experiment results showed that people are more susceptible to health misinformation if they have less education and health literacy, less trust in public health systems, and are more inclined to accept alternative medicines. In addition, they found that these hypotheses are not mutually exclusive.

% \cy{\cite{van2022misinformation} [nature medicine]}\hw{The survey mentioned the psychological factors for people being susceptible to misinformation, I added them here}

% \cy{\cite{traberg2022psychological}Psychological inoculation against misinformation}
% 
% \cy{\cite{cinelli2022conspiracy} Conspiracy theories and social media platforms\cite{van2022psychological}Psychological Benefits of Believing Conspiracy Theories}
% 
% \cy{The process of psychological inoculation against misinformation~\cite{van2022misinformation}}
% 
% 
% \cy{\cite{delmastro2022depression} The spread of misinformation and conspiracy theories related to COVID‐19 has represented one of the several undesirable effects of the current pandemic. In understanding why people can be more or less at risk to believe in misinformation, emotional distress and education could play a crucial role. The present study aims to analyze the relationship among depressive symptoms, education, and beliefs in misinformation about COVID-19 during the early phase of the pandemic.}

\noindent \textbf{Sociological Drivers of Health Misinformation.}
% \ks{we may find some research on the effect of people's vulnerability to misinformation due to the lack of SCIENTIFIC and public HEALTH knowledge; you may also check other surveys listed in section 6 such as \cite{cacciatore2021misinformation} } 
One leading incentive for the spread of health misinformation is the profit from selling medical products online \cite{hu2002product, warner2022online}. With the increasingly accessible health information taking over the internet, the overall quality and credibility of online health information have plummeted. By studying the returned results for different skin condition searches on top internet search engines, Hu et al. \cite{hu2002product} found that product-related sites outnumbered educational sites. Their results showed that consumers are at risk of being negatively influenced by misinformation and buying ineffective products. Another study conducted by Warner et al. \cite{warner2022online} showed a similar conclusion; by analyzing the search results of cancer nutrition information on Pinterest, they found that almost half of the content sites were for profit.
Another contributing sociological driver is the lack of criminal treatment against bad players on the internet, Tsirintani \cite{tsirintani2021fake} discovered that due to the anonymous nature of the internet, it is easy for perpetrators to conceal their identity and spread health misinformation, like creating bot accounts to retweet misinformation. Therefore, it adds difficulty for public officials to pursue cybercrime charges against offenders. In addition, the cross-border nature of cybercrime creates problems with jurisdiction boundaries. International consensus has not been established on the protocol for dealing with the globalization of criminal behaviors on the Internet.
Moreover, Zhou et al. \cite{zhou2021characterizing} provided an interesting perspective that treats online health misinformation as providing false social support for individuals to spread misinformation. Specifically, Liang et al. \cite{liang2011drives} stated that social support includes informational support and emotional support. Therefore, misinformation can be seen as false social support for individuals during health emergencies where people are eager to seek information regarding how to prevent or cure the disease and emotional support from others.

% \cy{the dissemination mechanism of health-related misinformation about COVID-19.~\cite{zhou2021characterizing}Health caution and advice, health help seeking misinformation, and emotional support play a significantly role in health misinformation spreading.
%  Misinformation ambiguity strengthens the effect of health caution and advice misinformation and weakens the effects of health help seeking misinformation and emotional support.
%  Misinformation richness strengthens the effect of health caution and advice misinformation and weakens the effect of emotional support.} \hw{added social support theory}

% \cy{a two-stage G-SCNDR model to simulate the spread of online rumor.\cite{wang2021rumor}}

\subsection{Effects of Health Misinformation}
% In 2021, U.S. surgeon general Dr. Vivek Murthy released a statement that called for a public effort to combat health misinformation. 
A recent statement released by the US surgeon general called for a public effort to combat health misinformation \cite{office2021confronting}. This statement showed that health misinformation poses unprecedented threats to our society on multiple levels and brings new challenges to the current research on misinformation. In this subsection, we characterize the negative consequences of health misinformation based on its level of impact. Specifically, we describe how health misinformation causes public confusion about scientific facts, negatively impacts public physical and mental health, and destruct social stability.

\noindent \textbf{Devaluation of Scientific Facts}
Unlike misinformation in the political domain where consumers of fake news in echo chambers have a hard time trusting information other than their favored existing narratives \cite{shu2017fake}, science is supposed to be a common ground where people with opposite ideological bents can agree on facts regardless of their opinions. Nonetheless, the public is losing its faith in scientific facts due to the explosive amount of untruthful scientific articles that were manipulated to create health misinformation. By studying the correlation between scientific quality and viewer engagement among prostate cancer videos on YouTube, Loeb et al. \cite{loeb2019dissemination} discovered that people are more inclined to watch poor-quality or misleading videos rather than high-quality videos based on scientific research evidence. In a separate study, Loeb et al. \cite{loeb2021quality} found that nearly 70\% of bladder cancer videos on YouTube are deemed to be misleading.

Unfortunately, this issue can be magnified by the abruption of a global public health crisis such as COVID.
% has been magnified by the abruption of a global pandemic such as COVID. 
%West and Bergstorm \cite{west2021misinformation} studied specific problems that are unique to misinformation in and about science such as publication bias, citation misdirection, predatory publishing, and filter bubbles. \ks{They found out that ...}
Scheufele et al. \cite{scheufele2021misinformation} studied misinformation about science in the context of public responses to COVID-19 informational environment. They proposed broad principles for communicating science with the public to avoid the spread of misinformation. Specifically, these principles are ``the need to tailor efforts toward clearly defined goals for communicating science, the importance of theory-based and practice-informed hypotheses for evaluating approaches to communicating, the need for metrics when assessing effectiveness, and the importance of partnerships between researchers and practitioners for generating needed evidence.''

\noindent \textbf{Damages to Public Health}
Besides the distortion of scientific facts, health misinformation causes harm to the physical and mental health of the general public \cite{lurie2022covid, verma2022examining}. For example, one critical case is the vaccine hesitancy caused by health misinformation during the COVID-19 pandemic. Lurie et al. \cite{lurie2022covid} studied the prevalence of COVID-19 misinformation in traditional news media. They analyzed the top vaccination misinformation themes and their trends. Their results indicated that although COVID-19 vaccine misinformation in traditional media is uncommon, they have the capacity to influence a large number of readers and affect their decision-making in getting the vaccine. On the contrary, research showed that misinformation regarding COVID-19 on social media is strongly associated with the public's willingness to get vaccinated \cite{xin2022roles, wang2022matter, gasteiger2022characteristics,samal2021impact}. 
% Concretely, Jiang et al. \cite{jiang2022covaxnet} created \verb|CoVaxNet|, a comprehensive online-offline data repository for research such as COVID-19 vaccine misinformation and stance detection.\ks{We may talk about CoVaxNet in the dataset section}
% \cy{add Zhou et al.'s work here}
% Besides the research from the societal perspective, there are also some research works that studied the causal effect of health misinformation on public health. Zhou et al.~\cite{zhang2022counterfactual} proposed a neural temporal point process model to conduct an unbiased estimation on the Individual Treatment Effect (ITE) of health misinformation from a data-driven perspective. \ks{canyu, do not add it here, figure our where to add}
Mental health is also exploited by the spread of COVID-19 misinformation. Research showed that the consumption of online misinformation takes a toll on its consumers' mental health. Verma et al. \cite{verma2022examining} studied the relationship between the consumption of misinformation on social media and its impact on mental health. By conducting a large-scale observational study, they discovered that there is a positive correlation between sharing COVID misinformation and experiencing anxiety.

% \cy{vaccine hesitancy\cite{rathje2022social}}
% 
% \cy{\cite{lee2022misinformation} Misinformation of COVID-19 vaccines and vaccine hesitancy\cite{pierri2022online}Online misinformation is linked to early COVID-19 vaccination hesitancy and refusal}

\noindent \textbf{Social Polarization}
Last but not least, health misinformation polarizes society when marry with political ideology. Soares et al. found that COVID-19 misinformation in Brazil was heavily politically motivated, and benefited far-right views\cite{soares2021research}. Yang et al. \cite{9679888} analyzed misinformation on social media platforms related to COVID-19 during the 2020 US election and found a strong correlation between the two events. Apart from the misinformation campaigns motivated by political elections, these campaigns are also used by foreign countries to push their own geopolitical interests. Specifically, Caniglia \cite{caniglia2020signs} discovered that during the COVID-19 outbreak, Italy became a central battleground of misinformation by state-sponsored operations between the United States and China, as well as between the EU and Russia.

Conversely, this social polarization effect reinforces people's political ideology and further consolidates their information bubble. By leveraging semantic features and network structures of Tweets regarding COVID-19, Jian et al. \cite{jiang2021social} modeled user polarity in terms of political preferences. They found that right-leaning users were noticeably more inclined to produce and consume COVID-19 misinformation. Additionally, their empirical evidence suggested that although echo chambers exist in both right- and left-leaning communities, the conservative political community was more connected with their echo chamber and more likely to reject information from the rest.

% \ks{The previous discuss is good. This section can be expanded a bit to talk about the other direction: polarization also affects misinformation consumption}

% \noindent \textbf{Causal Influence Modeling}
% \cite{zhang2022counterfactual} 

% \cy{discussion on how social media affects health information}
% 
% social media impact on mental health care
% \cite{de2013predicting}

%\noindent \textbf{Health Disparity}
%\cite{chang2021mobility}
%Reinforce the existing health disparity

% \cy{\textbf{misinformation besides covid}
% 
% \cite{elsherief2021characterizing}
% Misinformation Relating to Medication for Opioid Use Disorder
% 
% 
% \cite{chou2018addressing} Addressing the Challenges of Cancer Misinformation on Social Media
% \cite{grimes2022struggle}Struggle against Cancer Misinformation
% \cite{walsh2020social,johnson2022cancer}
% }

% \cy{\noindent \textbf{Escalation of Health Disparities}}

% \cy{\noindent \textbf{Amplification of Public Distrust}}
% \cy{In sub-Saharan Africa, researchers have argued that this skepticism extends to “distrust of philanthropic institutions, distrust of developed nations, and even distrust of leaders in their own respective countries” ~\cite{nwankwo2020topic}}

\subsection{Influence Strategies of Health Misinformation}
In previous subsections, we characterized the causes and effects of health misinformation with interdisciplinary theories. We further explore how misinformation campaigns use various influence strategies and dissemination patterns to spread health misinformation online. The insights provided in this section will advance our understanding of various means of health misinformation influence and guide its detection and intervention.

\noindent 
% \textbf{Racial Targeting}
\textbf{Targeting Populations with  Health Disparities}
One important strategy utilized by misinformation campaigns is targeting populations with health disparities, such as certain racial and religious groups. Diamond et al. \cite{10.1145/3491102.3501892} found that misinformation campaigns on the COVID-19 vaccine are inflammatory and purposely target certain racial groups. Through a primarily qualitative study, the authors found that such racial targeting is intertwined with issues such as medical racism and vaccine hesitancy. Similarly, Loobma et al. \cite{loomba2021measuring} discovered that different sociodemographic groups are impacted differently when exposed to misinformation and showed a strong association between misinformation and the decreased vaccination intention in the UK and the USA. To understand the reasoning for racial targeting, Austin et al. \cite{austin2021covid} found that people of color are disproportionately targeted by misinformation campaigns due to the reported discrepancy in media literacy compared to their white counterparts. Their results suggested that people with less media literacy have more acceptance of COVID-19 misinformation and therefore are easily victimized by misinformation.

Certain religious groups were targeted during the COVID-19 pandemic \cite{druckman2021role}. Kanozia and Arya \cite{kanozia2021fake} found that certain racial groups were purposely targeted by misinformation campaigns in India, Pakistan, and Bangladesh. For example, false claims were spreading among Hindu communities that cow meat was being used when developing COVID-19 vaccines. Similar claims have also been made targeting Muslim communities stating that the vaccine can be traced to cow meat. Furthermore, Al-Zaman \cite{al2020politics} found that most of the COVID-19 misinformation in India targeting Muslims is motivated by Islamophobia, and misinformation targeting Hinduism is related to the contemporary political ideologies of India.

% \ks{racial targeting is too narrow. Find some examples of disparity issue other than race. @Canyu, provide some papers here}

\noindent \textbf{Leverage Scientific Uncertainty}
When encountering newly emerged diseases such as COVID-19, there exists a great number of scientific uncertainties. These scientific uncertainly could be adversely leveraged by misinformation campaigns. One example is the mask policy in the U.S. at the beginning of the outbreak, the CDC did not explicitly claim the effectiveness of masks against COVID due to scientific uncertainty. This ambivalence was later exploited to fabricate misinformation about masks being a political tool to control people \cite{he2021people}. 
Granter and Papke \cite{granter2018medical} found that the prevalence of untruthful scientific documents such as unqualified citations of retracted articles, and studies that have been contradicted by studies possessing more evidence have been used to promote pseudo-science to benefit the spread of health misinformation. Acher and Chaiet \cite{acker2020weaponization} studied a newly emerged misinformation tactic linked to the COVID-19 pandemic where fraudsters targeted using archived health misinformation to propagate false beliefs on platforms such as Facebook and Twitter. Their research showed that such archived web sources can be difficult for automated systems to moderate, resulting in a longer and wider spread on social media.

% \cy{\cite{safarnejad2020contrasting} Structural Dissemination Patterns of Health Misinformation, \cite{nsoesie2020identifying}Transmission patterns}

% \subsection{Influence Pathways of Health Misinformation}
\subsection{Dissemination Patterns of Health Misinformation}
% \ks{use specific social media platforms as the subtitles seems to loosely related to pathways. You should clearly define pathways and come up with different subtitles in the following, use facebook, twitter, youtube, tiktok, etc only as examples. Rewrite.}

% \cy{\noindent \textbf{Private Influence Pathways/Channels} facebook, snap, characteristics: mislead the audience more easily, smaller impact scope}

% \cy{\noindent \textbf{Public Influence Pathways/Channels} twitter, youtube, tiktok, characteristics: mislead the audience harder, larger impact scope}

% Research \cite{prabhu2021capitol} has shown that the misinformation that incited the January 6 capitol riot started on Parler, a right-wing social media platform, and then quickly spread across mainstream social media such as Twitter and TikTok. These dynamic means by which misinformation flows from less reputable information sources to popular media platforms are caused by the distinct dissemination pattern of health misinformation. 
% With the increasing threats caused by offline consequences of online misinformation, understanding and modeling influence pathways \cite{kettler} have been of great interest to U.S. national security agencies such as DARPA.

In this subsection, we discuss the characteristics of dissemination patterns for health misinformation from two perspectives: propagation structure and influence channel.

\noindent \textbf{Propagation structure}
Health misinformation is transmitted at both a high level that follows certain topics (COVID-19 vaccine is against Islamic law \cite{mardian2021sharia}) and a low level that targets specific populations (COVID-19 vaccine is developed using pork \cite{kanozia2021fake}). One example of studying dissemination patterns at a high level is topic modeling. Liu et al. \cite{liu2020health} performed topic modeling on news reports during the early stage of the COVID-19 outbreak in China. They found that the top three most popular topics were ``prevention and control procedures, medical treatment and research, and global or local social and economic influences.'' Similarly, Jamison et al. \cite{jamison2020adapting} extended the typology of vaccine misinformation on Twitter by using manual content analysis and Latent Dirichlet Allocation (LDA). Additionally, Safarnejad et al. \cite{safarnejad2020contrasting} investigated the dissemination patterns of Zika misinformation on Twitter in 2016. Specifically, they reconstructed retweeting networks for both misinformation and real-information tweets. By comparing the network structure metrics between the two networks, they found that misinformation network structures were generally more complex than real-information networks.

%Nsoesie and Oladeji \cite{nsoesie2020identifying} 

% \noindent \textbf{Distinct Content} 
% %\ks{this part needs to be extended}
% Health misinformation propagates based on its unique content. The way health misinformation disseminates on social media is very different than political misinformation. As discussed in the previous subsection, health misinformation relies on manipulating scientific uncertainty or prompting pseudo-scientific evidence to mislead its targets. Additionally, health is a smaller domain compared to other misinformation domains such as politics, and is less studied among the misinformation detection community. Finally, debunking health misinformation requires specialized knowledge of public health or biomedicine, which adds challenges to fact-checking at a large scale. Consequently, the distinct content of health misinformation makes its influence extremely misleading, highly transmissible, and very hard to fact-check.

\noindent \textbf{Influence Channel}
Health misinformation spread across a variety of social media platforms. Sharma et al. \cite{sharma2017zika} discovered that misguided video posts about the Zika virus on Facebook were far more popular than the posts dispersing accurate public health information about the disease; Suarez-Lledo and Alvarez-Galvez \cite{suarez2021prevalence} found that misinformation regarding topics like vaccines, eating disorder, and medical treatments are prevalent on Twitter; Li et al. \cite{li2022youtube} found that YouTube is the primary source of misinformation on COVID-19 vaccination. More importantly, studies \cite{baumel2021dissemination, zenone2021tiktok} have shown that TikTok is used as a medium to propagate health misinformation among adolescent users. The way TikTok continuously recommends new videos based on users' browsing history puts the users in a vulnerable situation where they could receive misleading information repeatedly. Despite the threat to public health among young people, health misinformation on TikTok has not been studied as widely as other social media platforms like Facebook and Twitter. In addition to the diverse mediums through which health misinformation is disseminated, studies have shown that health misinformation flows in a cross-platform manner. Ginossar et al. \cite{ginossar2022cross} discovered that anti-vaccination misinformation campaigns utilize a hybrid strategy of tweeting links of YouTube videos to promote misleading information.

\section{Health Misinformation Detection}\label{sec:detect}
In this section, we present the detailed techniques of detecting health misinformation. Generally, the detection models can be categorised into six classes: capturing health content, exploiting user engagements, modeling social structures, incorporating biomedical knowledge, connecting multiple modalities and leveraging other domains.
% \ks{TODO}

% We will 
% first briefly  introduce traditional computational methods that are applied in common misinformation detection, and then 
% describe in details of detection methods focused in the health domain. We also discuss the unique points of health misinformation detection.

% \subsection{Traditional Misinformation Detection}
% \subsection{Health Misinformation Detection}

% \usepackage{tikz}
\usetikzlibrary{arrows.meta,shapes,positioning,shadows,trees}

\tikzset{
    basic/.style  = {draw, text width=5.3cm,  text height=0.3cm, drop shadow, font=\sffamily, rectangle},
    root/.style   = {basic, rounded corners=3pt, thin, align=center, fill=yellow!60},
    onode/.style = {basic, thin, align=center, fill=yellow!30,text width=5.7cm,},
    edge from parent/.style={->, >={latex}, draw=black, edge from parent fork right}
}

% \begin{document}
\begin{figure*}[!htbp]
    \centering
\begin{tikzpicture}[%
    grow=right,
    anchor=west,
    growth parent anchor=east,
    parent anchor=east,
    level 1/.style={sibling distance=1.1cm},
    level distance=1cm]

    \node[root] (root) {Health Misinformation Detection}
    child {node[onode] (c1) {Leveraging Other Domains}}
    child {node[onode] (c2) {Connecting Multiple Modalities}}
    child {node[onode] (c3) {Incorporating Biomedical Knowledge}}
    child {node[onode] (c4) {Modeling Social Structure}}
    child {node[onode] (c5) {Exploiting User Engagements}}
    child {node[onode] (c6) {Capturing Health Content}}
;
\end{tikzpicture}
    \vspace{0.1cm}
    \caption{Overview of Health Misinformation Detection Methods}
    \vspace{-0.1cm}
    \label{Detection}
\end{figure*}
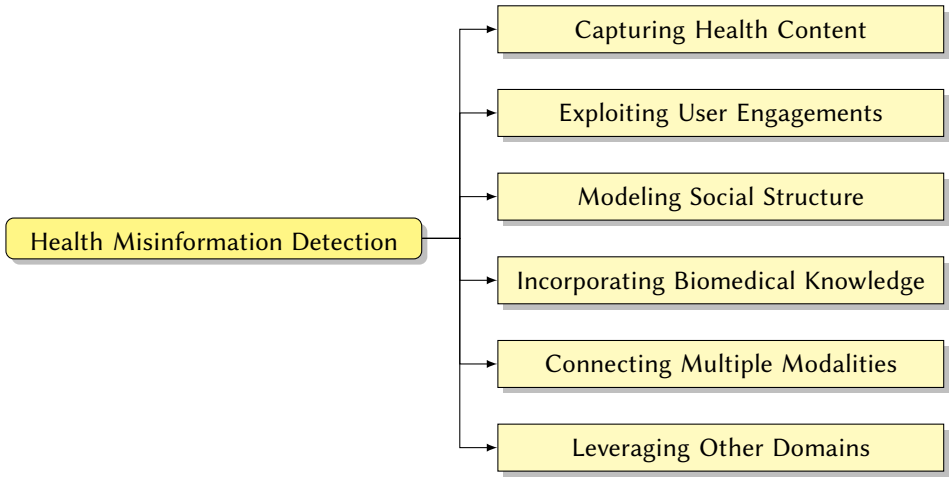

\subsection{Capturing Health Content}
Health misinformation usually contains opinionated or sensational linguistic features as they are intentionally created to misguide the audience~\cite{ wang2019systematic,suarez2021prevalence}. These linguistic features can be exploited for detection. For example, considering the false claim ``The shot is not just a vaccine but the government's conspiracy.'' This exaggerative language style indicates that it is likely a piece of health misinformation~\cite{sanaullah2022applications}.
% \ks{add references}.

Extensive research has been done on exploiting the content semantics for identifying health misinformation~\cite{tashtoush2022deep,kumari2021debunking,wan2022data, aphinyanaphongs2007text,li2021exploring,10.1145/3487553.3524662,Chen2022UsingFC,khan2022detecting, 9599970,AlAhmad2021AnEF,gundapu2021transformer,joy2022comparative,Pavlov2022COVID19FN,Hande2021EvaluatingPT,Chen2022UsingDL, Abdelminaam2021CoAIDDEEPAO}. Some papers utilize classical machine learning algorithms to detect  health misinformation. For example, Shushkevich et al.~\cite{9599970} utilized word frequency in datasets to convert text into vectors. Then they adopted Logistic Regression (LR)~\cite{cox1958regression}, Support Vectors Machine (SVM)~\cite{cortes1995support}, Gradient
Boosting (GB)~\cite{friedman2001greedy}, Random Forest (RF)~\cite{ho1995random} to detect COVID-19 fake news. Al-Ahmad et al.~\cite{AlAhmad2021AnEF} proposed an evolutionary detection approach for health misinformation by leveraging Particle Swarm Optimization (PSO)~\cite{wang2021particle}, Genetic Algorithm (GA)~\cite{deng2021improved} and Salp Swarm Algorithm (SSA)~\cite{khurma2021salp}.
% conventional methods \cite{AlAhmad2021AnEF} \ks{Some papers leverage traditionally language processing and extract features ... For example, paper xx found out that "xxx" are strongly indicative phases for detection misinformation of cancer/diabetes..}

With the development of deep learning for natural language processing, recent works leverage deep neural networks to detect health misinformation. 
For example, Tashtoush et al.~\cite{tashtoush2022deep} investigated several typical machine learning models (e.g., Long Short-Term Memory (LSTM)~\cite{hochreiter1997long}, Convolutional Neural Network (CNN)~\cite{kalchbrenner2014convolutional}, and Bidirectional Encoder Representations from Transformers (BERT)~\cite{devlin2018bert}) to capture the linguistic cues from the content. In addition, Kumari et al.~\cite{kumari2021debunking} proposed a self-ensemble SEIBERT (Scientific BERT~\cite{DBLP:conf/emnlp/BeltagyLC19}) model that utilizes domain-specific word embedding for detecting health information. Chen et al~\cite{Chen2022UsingFC} proposed to combine deep neural network models with fuzzy logic to detect COVID-19 misinformation. To improve the computation efficiency while maintaining high prediction performances, they used fuzzy clustering to reduce the redundant features in the model.
% the detection performances of multiple basic deep neural networks including Long Short-Term Memory (LSTM), Bi-directional LSTM (BiLSTM), Convolutional Neural Network (CNN) solely based on the news contents. 
% Since BERT (Bidirectional Encoder Representations from Transformers) has shown powerful in various natural language processing tasks due to pretraining on a large corpus, Kumari et al.~\cite{kumari2021debunking} proposed a self-ensemble SCIBERT (Scientific BERT) based model that utilizes domain-specific word embedding for detecting health information.
% \cy{classical \& neural network methods}
% \cy{compared with non-health fake news detection, what's the difference of health fake news detection}

\subsection{Exploiting User Engagements} The user engagements of health misinformation is about the surrounding context where users interact with the misinformation on social media. For example, in Twitter, the engagements include various common actions including tweeting, retweeting, commenting, clicking, liking and disliking~\cite{di2022assessing,vyas2021proliferation}. Since health misinformation tends to be controversial and confusing, the social engagements around it are likely to include more intense sentiments and polarized responses, which can be valuable features for automatic detection~\cite{darwish2022survey}. For example, considering a misleading claim ``The vaccine will make you die tomorrow if you take it today'' on Twitter. The comments can be useful to indicate its falsehood such as ``You are ridiculous'', ``That's crazy!'' 

Recently, a significant amount of efforts have been made to incorporate various user engagements information for detecting health misinformation online~\cite{zhao2021detecting,di2022health,ghenai2018fake}. These papers generally demonstrate the need and effectiveness of user engagements to improve the detection of health misinformation. 
Ghenai et al.~\cite{ghenai2018fake} examined the users' behavior about promoting false cancer treatments on Twitter and then proposed a user-centric model for detecting users prone to propagate health misinformation by extracting user behavior features including users’ attitudes, writing styles, and post sentiments on social media. In addition,
Zhao et al.~\cite{zhao2021detecting} further studied four types of user engagement features including linguistic features, topic features, sentiment features, and behavioral features for detecting misinformation in online health communities. They found out that social behavioral features are most helpful for the machine learning based detection models. 
% In additiona, 
% Sotto et al.~\cite{di2022health} analyzed six types of health misinformation features that consist of textual representation features, linguistic-stylistic features, linguistic-emotional features, linguistic-medical features, propagation-network features, and user-profile features.
% \\

\subsection{Modeling Social Structure} The social structure of health misinformation refers to the network structures which the misinformation interact with on social media. The scope of social structure is beyond user engagements and embraces various concepts including message propagation trajectories, user-user and user-post networking. Propagation trajectories describe the path of misinformation circulation on the social networks. User-user and user-post networking describe social networking where users interact with users and users interact with posts accordingly.
% includes its social engagements, propagation trajectories and user connections on social media. Social engagements include the tweet and retweet behaviours of the news, which reflect the users' stance towards the news content. Propagation trajectories describe the path and scope of misinformation circulation on the social networks. User connections refer to the user profiles who interact with the news content in various ways including commenting, clicking, liking and disliking.

The social structure of health misinformation dissemination is valuable for differentiating health misinformation from real information. Safarnejad et al.~\cite{safarnejad2020contrasting} studied the  real and false posts about Zika Virus disseminated on Twitter from a dynamic network perspective and found out that they have distinct social network structures. For example, they discovered that the propagation of health misinformation tends to form more local clusters and has a larger MOD score than real information (MOD is a local-level network metric measuring the likelihood of dividing a graph network into potential clusters).  Their findings indicate that the unique propagation patterns of health misinformation can be utilized for developing more accurate detectors. 

The social structure of health misinformation can be captured and represented as graphs. Thus, some existing works~\cite{karnyoto2022augmentation,dhanasekaran2021sompsnet,hamid2020fake,10.1145/3485447.3512163,paraschiv2021unified,10.1145/3511808.3557394}  explored utilizing Graph Neural Network (GNN) models for the detection. For instance, Prasannakumaran et al. ~\cite{dhanasekaran2021sompsnet} proposed a GNN-based model named SOMPS-Net which jointly models the social engagements, publisher details, and article statistics to achieve high detection performances. Karnyoto et al.~\cite{karnyoto2022augmentation} investigated multiple graph neural network models including Graph Attention Network (GAT)~\cite{velickovic2017graph}, Graph Convolutional Network (GCN)~\cite{yao2019graph}, and GraphSAGE~\cite{hamilton2017inductive} with several augmentation strategies in COVID-19 misinformation detection. To tackle the challenge of heterogeneity in social networks,
Min et al.~\cite{10.1145/3485447.3512163} formulated the detection as a heterogeneous graph classification task and proposed to model the post-post, user-user and post-user interactions on social network with a divide-and-conquer strategy. In addition, Cui et al.~\cite{10.1145/3511808.3557394} utilized a meta-path on the heterogeneous social network to extract multi-level context information (news publishers and engaged users) and  temporal information of user interactions. Paraschiv et al.~\cite{paraschiv2021unified} pointed out that treating user information and network information in isolation could not fully characterize the manifestation of misinformation and proposed to model the user-based, network-based, content-based features of health misinformation with a unified meta-graph structure.

% \\
\subsection{Incorporating Biomedical Knowledge} 
% \ks{two aspects: matching-based and knowledge-enhanced}
Generally, the biomedical knowledge incorporated into health misinformation detection can be categorized into \textit{graph-level} and \textit{text-level} knowledge, which refer to \textit{biomedical knowledge graphs} and \textit{biomedical evidential texts} accordingly. 
Guiding health misinformation detectors with biomedical knowledge has shown to be beneficial from the following perspectives: (1) biomedical knowledge can be more reliable auxiliary information than online content because they are built on convincing research papers and medical reports. For example, for medical entities Actonel and Hypocalcemia, although it is difficult for normal people to judge whether or not Actonel  can heal Hypocalcemia, we can retrieve the related knowledge from biomedical knowledge graphs easily; (2) bringing biomedical knowledge can enhance the explainability and trustworthiness of health misinformation detection. Because there may be many medical terminologies in the misinformation, it is hard to convince normal users who have little health expertise and let them accept the detection results without offering understandable explanations~\cite{10.1145/3394486.3403092,10.1145/3492855,abu-salih2022healthcare}.

The \textit{biomedical knowledge graphs} are constructed by health experts from peer-reviewed research papers and authoritative reports. Generally, they contain the health entities collected from biomedical literature and the positive/negative  relations (e.g., Heal/DoesNotHeal) between entities. For example, for entities Calcium Chloride and Hypocalcemia, the triplet (Calcium Chloride, Heal, Hypocalcemia) has a positive relation. For entities Actonel and Hypocalcemia, the triplet (Actonel, DoesNotHeal, Hypocalcemia) has a negative relation~\cite{10.1145/3394486.3403092}.
There are some emerging works that studied exploiting biomedical knowledge graphs for health misinformation detection~\cite{10.1145/3394486.3403092,10.1145/3492855,9812879,koloski2021knowledge,mayank2021deapfaked,10.1145/3437963.3441828}.
Cui et al.~\cite{10.1145/3394486.3403092} proposed a medical knowledge graph guided model named DETERRENT for identifying health misinformation. Specifically, they first constructed a unified graph based on an Article-Entity Bipartite Graph and a Medical Knowledge Graph. Then they utilized an Information Propagation Net to propagate the knowledge from knowledge graphs to news articles. Finally, they adopted a Graph Attention Network (GAT) to aggregate the information and make the prediction. 
Extensive experiments on two real-world misinformation datasets on diabetes and cancer demonstrated the effectiveness of biomedical knowledge graph guidance. 
% \ks{why describing paper 23 with so many details and not talking about 49, 84 at all?? Add some content for 49, 84 (do as what I told you here)}
% Kou et al.~\cite{10.1145/3492855}
In addition, Koloski et al.~\cite{koloski2021knowledge} utilized knowledge graphs to enrich the text representations for COVID-19 misinformation detection.
Mayank et al.~\cite{mayank2021deapfaked} proposed to combine the encoding of texts and knowledge graphs and discovered that they have a complementary advantage on health misinformation detection. To tackle the challenge that current knowledge graphs lack COVID-19-specific knowledge facts, Kou et al.~\cite{10.1145/3492855} constructed a crowdsource knowledge graph with the collaborative efforts of experts and non-expert workers. Then, they proposed a duo hierarchical attention-based graph neural network model HC-COVID that explicitly integrates the knowledge graph facts contributed by workers with different levels of expertise. Vedula et al.~\cite{10.1145/3437963.3441828} proposed to employ both the unstructured textual context and  structured knowledge graph to detect the veracity of health claims and generate human-understandable natural language explanations.

% \cite{}
% \subsubsection{Bring Health Experts in the Loop}

% \subsection{Accessing Crowd Wisdom}

% \cite{ghenai2017catching,kou2021fakesens,10.1145/3492855, kou2022crowd,9812879}

% In \cite{10.1145/3492855}, the authors proposed to construct a crowdsource knowledge graph with the collaborative efforts of experts and non-expert workers. The challenge is how to effectively coordinate the collaboration and then leverage the constructed knowledge graph. Thus, they proposed a duo hierarchical attention based graph neural network model HC-COVID that explicitly integrates the knowledge graph facts contributed by workers with different level of expertise. 

The \textit{biomedical evidential texts} is another major kind of external source to check the veracity of health-related claims on social media.
% False health-related claims have a large proportion of all health misinformation on social media\cite{di2022health}. Utilizing external evidence is an effective way to check the veracity of health-related claims. 
False medical claims have made up a large proportion of all health misinformation on social media\cite{di2022health}.
Different from judging the truthfulness of claims in other domains such as politics, it is vital to utilize up-to-date external biomedical evidence to verify a medical claim because novel research might change or overturn the existing views in biomedicine~\cite{whrl2022entitybased}. Fact-checking of health-related claims is typically a two-step process: evidence retrieval and verdict prediction~\cite{guo2022survey,DBLP:journals/tacl/GuoSV22}. The verdict then determined whether or not the evidence supported the claim. Utilizing biomedical evidential texts to help detect health-related claims has attracted increasing attention~\cite{aranacatania2022natural,saakyan2021covidfact,whrl2022entitybased,wadden2021multivers,wadden2022scifactopen,pradeep-etal-2021-scientific,li2021paragraph,zhang2021abstract,Sarrouti2021EvidencebasedFO,zeng2021qmulsds,kotonya2020explainable}. 
% \ks{the differences between this section and the biomedical knowledge is not clear. May be combining them}
% \cy{I change 3.4 to Biomedical Knowledge Graphs}
For example, Arana-Catania et al.~\cite{aranacatania2022natural} first built a dataset containing heterogeneous claims on Covid and then investigated multiple natural language inference methods  for claim veracity assessment. Rather than employing human annotators, Saakyan et al.~\cite{saakyan2021covidfact} utilized automatic methods to detect true claims, source articles, and counter-claims to construct a medical claim dataset and provided a strong claim verification baseline using RoBERTa-large~\cite{liu2019roberta}. Wührl et al.~\cite{whrl2022entitybased}  proposed to extract entity-based claim representations from user-generated medical contents online and showed that this reformulation improved multiple fact-checking models. Wadden et al.~\cite{wadden2021multivers} incorporated more relevant information for health claim validation via modeling the full-document context and designed a multi-task modeling approach for zero/few-shot scenarios. Some research works also leveraged biomedical evidential texts to provide human-comprehensible explanations for veracity prediction. For example, Kotonya et al.~\cite{kotonya2020explainable} adopted extractive-abstractive summarization models to develop explanations and found out that training veracity prediction model and explanation generation models together on in-domain data  can improve the detection performance of public false health claims.

% \cite{dougrezlewis2022phemeplus}
% \cite{aranacatania2022natural}
% \cite{saakyan2021covidfact}
% \cite{whrl2022entitybased}
% \cite{wadden2021multivers}
% \cite{wadden2022scifactopen}
% \cite{pradeep-etal-2021-scientific}
% \cite{li2021paragraph}
% \cite{zhang2021abstract}
% \cite{Sarrouti2021EvidencebasedFO}
% \cite{zeng2021qmulsds}
% \cite{liu-etal-2020-adapting-open}
% \cite{kotonya2020explainable}

% \ks{CC, you expand the papers first, and i will revise again}
% \\
% \textbf{challenges in Health Disinformation Detection}
\subsection{Connecting Multiple Modalities} 
Multimodal detection refers to detecting health misinformation based on multiple modalities including text, image, audio, video, table, etc. Different from most existing single-modal (e.g., text) based detection methods, multi-modal methods are increasingly desired due to the following reasons. \textit{First}, it is becoming important to detect health misinformation beyond the text modality. With the rapid growth of photo-sharing applications (e.g., Instagram, Flickr), video-sharing platforms (e.g., Tiktok, YouTube), and platforms that can share multimodal messages (e.g., Twitter, Facebook), the dissemination of health misinformation on social media is not limited to the format of text. For example, The Institute for Strategic Dialogue tracked 124 TikTok videos featuring anti-vaccine misinformation and found out that they have generated 339 thousand shares and more than 20 million views\footnote{https://www.isdglobal.org/digital\_dispatches/how-tiktok-sounds-are-used-to-fuel-anti-vaccine-fears/}. 
% More seriously, most of the users of TikTok are young people, who are more vulnerable to the covid inflection due to the relatively low vaccination rate in young people and minority groups\footnote{https://www.census.gov/library/stories/2021/12/who-are-the-adults-not-vaccinated-against-covid.html}.
\textit{Second}, most conventional health misinformation detectors focus on exploiting the textual or networking features on social media, which cannot be directly applied to the multimodal data~\cite{Guo2022ASR}. For example, the health misinformation disseminated on Instagram contains visual and textual modalities and that on TikTok contains visual, audio, and textual modalities. \textit{Third}, exploiting the coonection across modalities is valuable as they may contain complementary cues to better detect multimodal misinformation~\cite{10.1145/3485447.3512257}. Health misinformation usually abuses the association to mislead the audience. For example, consider a piece of vaccine misinformation that contains an image of shingles, showing the symptoms caused by the chickenpox virus, with the caption ``Side effects of COVID-19 vaccine''. The misinformation abuses the association of the image and caption to mislead the readers that the side effects of the COVID-19 vaccine include shingles.

There are increasing research works on exploring to utilize the multi-modality of social media posts for health misinformation detection~\cite{wang2020detecting,10.1145/3485447.3512257,zhou2021instavax,shang2021multimodal,silva2021embracing}. In \cite{wang2020detecting}, Wang et al. first collected a multimodal dataset from Instagram including both visual elements (e.g., photographs and posters) and textual elements (e.g., captions, words in images, and hashtags) to analyze the antivaccine messages. Then they proposed a three-branch attention mechanism to encode the information of hashtag, text, and image modalities separately. The three single-modal features are fused to make the final prediction. In addition, Shang et al.~\cite{10.1145/3485447.3512257} showed that the simply fusing the features from different modalities cannot effectively leverage their inner associations. For example, the association between the image of shingles and the caption ``Side effects of COVID-19 vaccine'' in one piece of misinformation can hardly be captured by simple modality fusion. Thus, they proposed a dual-generative scheme to explicitly model the deep cross-modal association and generate image-guided textual features and text-guided image features simultaneously for detection. Furthermore, they modeled the multimodal content and user comments as a Content-comment Graph to generate explanations. Zhou et al.~\cite{zhou2021instavax} build a benchmark for multi-modal misinformation detection, and demonstrate (1) the need of incorporating multi-modal information rather than single-modal, and (2) text context still play the majority role in the detection and visual information only brings limited benefits, which indicates there is still a large improvement space on how to better utilize the multiple modalities.
% collected a multimodal dataset named Insta-VAX that consists of 64,957 Instagram posts related to human vaccines, and then benchmarked three groups of models including \textit{Text-only Uni-modal Models}, \textit{Image-only Uni-modal Models} and \textit{Multi-modal Models}. 
% Based on extensive experiment results, they have several key findings. First, multi-modal models outperform the uni-modal models, which verifies the effectiveness of incorporating multiple modalities. Second, the text context still play the majority role in the detection and visual information only brings limited benefits, which indicates there is still a large improvement space on how to better utilize the multiple modalities.
For health misinformation disseminated on other channels such as video-sharing platforms, it is necessary to take multiple modalities into account. Shang et  al.~\cite{shang2021multimodal} first analyzed the video-based health misinformation on TikTok and pointed out that the algorithms on verifying video authenticity cannot be directly applied to detecting health misinformation in short videos. 
% Specifically, TikTok videos often have a maximum length of 60 seconds. Users can manipulate the short videos in multiple ways including superimposing captions, modifying the virtual background and  adding an anime filter. Thus, the health misinformation can manipulate the contents in multiple modalities to mislead the audience. For example, one misleading short video contains the visual content showing that a man has a magnetic strip on his arm, and the superimposed caption ``So here you're telling me that the shot isn't
% something more than a vaccine''. Although neither the visual part nor the caption part is misleading, the whole video jointly mislead the audience that vaccine can make our body magnetic. 
They proposed a caption-guided visual representation learning module to explicitly exploit the caption information in visual and audio frames, and a visual-speech co-attention module to model the correlation between visual frames and speech contents.

% \cite{silva2021embracing}

% \cite{10.1145/3485447.3512257}

% \cite{shang2021multimodal}

% ks{write and expand the paper!}
% \subsection{Trustworthy Detection}
% fairness~\cite{park2021presence}
% \cite{chang2022disparate}
% \cite{ayoub2021combat}
% \cite{10.1145/3492855} \cite{10.1145/3485447.3512257} explainableXX% \\% \\
% \cite{yue2022contrastive}
% \\
% \textbf{Fact Checking}

% \subsection{Other Methods}

% \textbf{Social Sensing Approach}

% \cite{ghenai2017catching}
% \cite{kou2021fakesens}\ks{write and expand the paper!}
% \\
% \\
% \subsection{Utilizing Visualization}

% \cite{kostkova2016vac}

\subsection{Leveraging Other Domains}

Misinformation has a wide dissemination over various domains including health, science, military, disasters, education, politics, finance, entertainment, and society due to the popularity of the internet~\cite{10.1145/3459637.3482139}. Leveraging other domains can bring a large advantage to health misinformation detection because the amount of misinformation data in health domain is relatively limited compared to the real-world data in various domains~\cite{liu-etal-2020-adapting-open,zhu2022memory}. Thus, it is promising to enhance the performances of  health misinformation detectors via jointly training on multiple domains or adapting the models trained on other domains.

A variety of approaches have been proposed to leverage knowledge in other domains to help health misinformation detection~\cite{nan2022improving,zhu2022memory,9671592,10.1145/3511808.3557263,suprem2022midas,10.1145/3532851,silva2021embracing,10.1145/3459637.3482139,Dhankar2021AnalysisOC,lin2022detect,liu-etal-2020-adapting-open,zeng2022unsupervised}. To tackle the data scarcity challenge of Covid fact checking, Liu et al.~\cite{liu-etal-2020-adapting-open} proposed rationale prediction based and masked language model based continuous training to adapt pre-trained language models to Covid domain. Zhu et al.~\cite{zhu2022memory} proposed a memory-guided multi-view multi-domain learning paradigm to solve the domain shift and domain labeling incompleteness challenges of training on multiple domains and improved the  performance of health misinformation detection. To solve the problem of early Covid misinformation detection, Yue et al.~\cite{10.1145/3511808.3557263} adopted contrastive domain adaptation and pseudo labeling with label correction to adapt misinformation detectors to the unseen Covid domain. Silva et al.~\cite{silva2021embracing} introduced a multimodal fake news detector that exploits both of cross-domain and domain-specific knowledge and achieved high performance for Covid fake news detection. In addition, Ding et al.~\cite{10.1145/3532851} proposed an end-to-end adversarial domain adaptation model named MetaDetector to enable event-level meta-knowledge transfer and enhanced the accuracy of health misinformation detection.

\section{Health Misinformation Intervention} \label{sec:intervention}
We have discussed the characteristics of health misinformation and covered the representative methods for detecting health misinformation. 
% Therefore, combating health misinformation requires a whole-of-society effort. 
In this section, we discuss intervention efforts needed from various players in our society. Specifically, we categorized interventions on three levels: government level, organization level, and individual level.

%% Credit: https://texample.net/media/tikz/examples/TEX/work-breakdown-structure.tex
\usetikzlibrary{arrows,shapes,positioning,shadows,trees}
\tikzset{
  basic/.style  = {draw, text width=6cm,  text height=0.3cm, drop shadow, font=\sffamily, rectangle},
  root/.style   = {basic, rounded corners=2pt, thin, align=center,
                   fill=green!60},
  level 2/.style = {basic, rounded corners=4pt, thin,align=center, fill=green!30,
                   text width=8em},
  level 3/.style = {basic, thin, align=left, fill=pink!30, text width=7em}
}

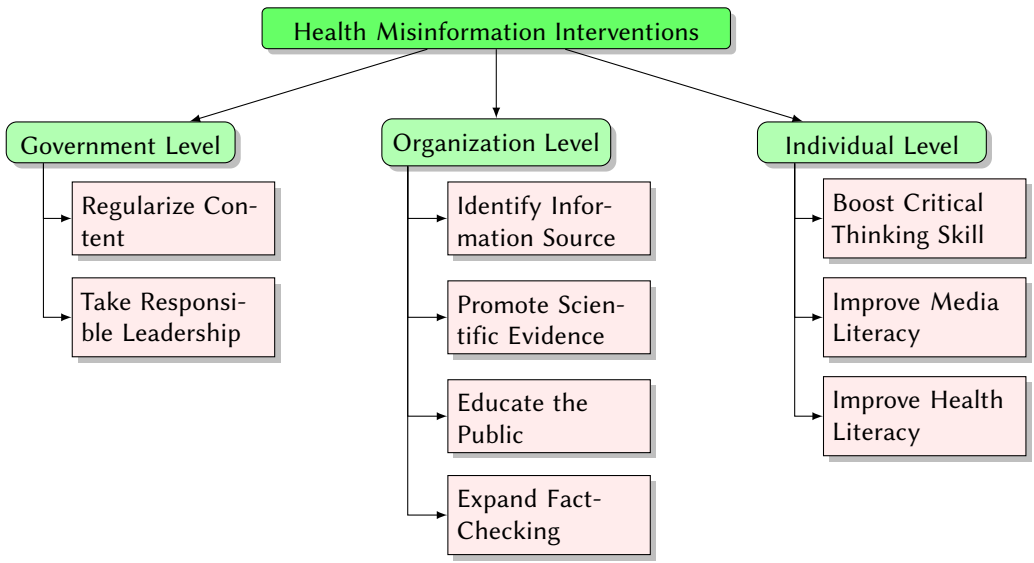
\begin{figure*}[!htbp]
    \centering
    \begin{tikzpicture}[
      level 1/.style={sibling distance=50mm},
      edge from parent/.style={->,draw},
      >=latex]
    
    % root of the the initial tree, level 1
    \node[root] {Health Misinformation Interventions}
    % The first level, as children of the initial tree
      child {node[level 2] (c1) {Government Level}}
      child {node[level 2] (c2) {Organization Level}}
      child {node[level 2] (c3) {Individual Level}};
    
    % The second level, relatively positioned nodes
    \begin{scope}[every node/.style={level 3}]
    \node [below of = c1, xshift=20pt] (c11) {Regularize Content};
    \node [below of = c11, node distance=13mm] (c12) {Take Responsible Leadership};
    
    \node [below of = c2, xshift=20pt] (c21) {Identify Information Source};
    \node [below of = c21, node distance=13mm] (c22) {Promote Scientific Evidence};
    \node [below of = c22, node distance=13mm] (c23) {Educate the Public};
    \node [below of = c23, node distance=13mm] (c24) {Expand Fact-Checking};
    
    \node [below of = c3, xshift=20pt] (c31) {Boost Critical Thinking Skill};
    \node [below of = c31, node distance=13mm] (c32) {Improve Media Literacy};
    \node [below of = c32, node distance=13mm] (c33) {Improve Health Literacy};
    \end{scope}
    
    % lines from each level 1 node to every one of its "children"
    \foreach \value in {1,...,2}
      \draw[->] (c1.195) |- (c1\value.west);
    
    \foreach \value in {1,...,4}
      \draw[->] (c2.195) |- (c2\value.west);
    
    \foreach \value in {1,...,3}
      \draw[->] (c3.195) |- (c3\value.west);
    \end{tikzpicture}

    \caption{Hierarchical Overview of Health Misinformation Interventions}
    \label{intervention}
\end{figure*}

\subsection{Government-Level Intervention}
%\ks{add transition}
In this subsection, we discuss the intervention efforts needed that government officials can provide. Specifically, we discuss the importance of content governance while preserving the right to freedom of speech and responsible public leadership.
%\subsubsection{Efforts from Policy Makers}

\noindent \textbf{Regularize Content} For legislators on capitol hill, the governance of online platforms is a direct approach to combat health misinformation online. However, how to balance the scale between preserving the right to freedom of speech and suppressing the spread of misinformation has been a long-standing issue. Specifically, Bowers and Zittrain \cite{bowers2020answering} studied this issue across three eras since the birth of the internet. They concluded that although it is almost impossible to make the trade-offs between accountability and freedom when it comes to content governance, it would be helpful to build transparency on how policy decisions are made and implemented. Similarly, Shattock \cite{shattock2021self} examined the impact Digital Services Act (DSA), a legislative measure to combat misinformation across the EU. They discovered that although DSA showed signs of improvement, it failed to address issues such as regulations of harmful but lawful content. Therefore, using legislation to regularize misinformation online is a double-edged sword, the government officials have to be very careful about not restricting freedom of expression. Pomeranz and Schwid \cite{pomeranz2021governmental} investigated the law passed in Zimbabwe that could prosecute people for up to 20 years in prison. Similarly, a law passed in the Philippines violate international law for suppressing expression freedom.

\noindent \textbf{Take Responsible Leadership} From a sociology standpoint, Antonakis \cite{antonakis2021leadership} explained why leadership makes a difference in the midst of a global pandemic. Government officials shape social values and affect the preferences or beliefs of their followers \cite{antonakis2022just}. Therefore public leaders need to prioritize promoting evidence-based scientific facts over the political agenda of their party or their own. In addition, the way public officials communicate their messages matters. Research has shown that charismatic public speeches make their messages have a broader influence on the public \cite{meslec2020effects}. More specifically, Antonakis et al. \cite{antonakis2016charisma} conceptualized charisma as ``signaling in a value-based, symbolic, and emotional manner.'' In conclusion, when facing a public health crisis, we need to make sure that office leaders are not only competent but also skilled communicators, charismatic and evidence-driven.

\subsection{Organization-Level Intervention}
%\ks{add transition}
In this subsection, we provide insights on the necessary interventions needed by different organizations. Concretely, we discuss the efforts needed by health organizations, social media platforms, and fact-checking organizations.

%\subsubsection{Efforts from Health Professionals and Organizations}
%\ks{add transition}

\noindent \textbf{Identify Information Source}
Media platforms are at the forefront of the battle against health misinformation. Although countermeasures have been taken, there exist flaws in strategies used by media platforms. Specifically, Courchesne et al. \cite{courchesne2021review} identified that 1) although there is abundant study regarding user-focused countermeasures, very little study has been done on countermeasures that impact the influence operations directly. 2) There exists a mismatch between the interventions taken by media platforms (e.g. algorithmic downranking, content moderation, deplatforming accounts, etc) and those studied by research communities.
Although media platforms have added the source of information to warn their users of potential misinformation. Research \cite{arnold2021source, dias2020emphasizing, nassetta2020state} have shown that such measures do not work effectively across the board and could even be counterproductive in some situations.
Social media platforms such as Twitter have implemented harm reduction tools such as source alerts (e.g. ``Determined by Twitter to be a Russian government account'') to combat misinformation. Arnold et al. \cite{arnold2021source} found that although source alerts can reduce the spread of misinformation online and offline, the partisanship of the user, the type of social media, and the specificity of the alert cause the effects to vary. Apart from the partisanship mentioned above, Nassetta and Gross \cite{nassetta2020state} found that whether or not the users notice the information source plays an important role in its effectiveness. Their experiment results showed that YouTube's funding labels can mitigate the effects of misinformation, but only when the users notice the labels. Finally, by conducting two survey experiments, Dias et al. \cite{dias2020emphasizing} pointed out that emphasizing source credibility may not be as effective as the intuition suggests, and could even have unintentional negative consequences.

\noindent \textbf{Promote Evidence-based Information} As discussed in Section \ref{char}, misinformation campaigns leverage ambiguous scientific facts to promote health misinformation. It is vital for the medical research community to publish evidence-based and reliable information to the public. However, evidence \cite{mheidly2020leveraging} showed that a mass of academic articles were published without rigorous peer review process and contained information that could be exploited by misinformation campaigns. To prevent health misinformation from wide dissemination, Mheidly and Fares \cite{mheidly2020leveraging} proposed ``Infodemic Response Checklist'' that calls for better media and health communication strategies against COVID-19 infodemic, including the health professionals. Similarly, Barr-Walker \cite{barr2021countering} called for efforts from health science librarians to  build partnerships with public health departments in order to provide evidence-based information about abortion and raise the public's awareness on abortion misinformation.

% \cy{scientists can help\cite{fleming2020coronavirus,diethelm2009denialism, farrell2019evidence}We recommend that expert organizations like the CDC immediately and personally rebut misinformation about health issues on social media.\cite{vraga2017using}}

\noindent \textbf{Educate the Public}
Research \cite{lazer2018science} has shown that education is a longer-run intervention approach to improve individual evaluation of the quality of information. Concretely, Osborne and Pimentel \cite{osborne2022science} discussed four ways that education can address scientific misinformation: ``adapting teacher training, developing curricular materials, revising standards and curricula, and improving assessment.'' With COVID-19 misinformation spreading on a rampage, it is important to educate the general public about scientific facts. Bahrami et al. \cite{bahrami2019counteracting} discussed the importance of educating the public with correct information regarding health issues and raising public awareness of health misinformation. Specifically, they suggest trained health workers share accessible health knowledge among the public, and raise public awareness about the prevalence of online health misinformation. On the other hand, Rubin \cite{rubin2022physicians} discussed the consequences of physicians sharing health misinformation with their patients. It not only puts patients' health in danger but also undermines public trust in health organizations. 

%\cy{\textbf{references from Science} \cite{osborne2022science,lazer2018science}}

\noindent \textbf{Expand Fact-checking}
Factuality is a key factor in the battle against health misinformation, which cannot be done without the effort from fact-checking efforts. However, fact-checking is labor-intensive and a scarce resource, especially during a global pandemic. By analyzing COVID-19 misinformation manually checked by fact-checkers, Simon et al. \cite{brennen2020types} found that fact-checking organizations had a hard time keeping up with the large amount of information circulating on social media. Therefore, we cannot solely rely on fact-checking organizations and need more ways for people to contribute their effort. Kim and Walker \cite{kim2020leveraging} found that many social media users correct misinformation when they encounter it. Based on this intuition, they implemented a strategy to leverage the efforts of volunteer fact-checkers and a social network to identify emerging COVID-19 misinformation. Alam et al. \cite{firoj_covid:icwsm21} took a holistic approach to support crowdsourcing annotation efforts. Specifically, they proposed an annotation schema and detailed annotation instructions that combine different players' perspectives in the fight against the infodemic. To address the shortage of fact-checkers, Miyazaki et al. \cite{miyazaki2022fake} examined the effect of spontaneous debunking from social media users to COVID-19 false information. They found that spontaneous debunking behavior is generally slower and less reliable than other responses.
%\cy{\cite{alam2020fighting,swire2019public}}

% \cite{brennen2020types} \ks{include this paper}

% a very related survey on health misinformation intervention
% \cy{\cite{janmohamed2021interventions, janmohamed2022interventions}}

\subsection{Individual-Level Intervention}
%\ks{add transition}
In this subsection, we discuss the actions that social media users can take to prevent themselves become the victim of online health misinformation.

\noindent \textbf{Boost Critical Thinking Skill}
Brashier et al. \cite{brashier2021timing} found that when people receive both misinformation and its correction, both information will retain, but the correction fades from the memory over time. Therefore, the timing of correction plays an important role in stopping the spread of misinformation. One technique to stop the consumption of misinformation at an early stage is by pausing to think about why a news headline is true or false. Fazio \cite{fazio2020pausing} conducted experiments where they ask the participants to explain why they rate the news headlines as real or fake. Their results suggested that forcing people to pause and think could reduce the spread of misinformation. Additionally, Pennycook et al. \cite{pennycook2021shifting} found that the current design of social media applications where users scroll quickly through a mixture of bipartisan and heavily opinionated content discourages people from reflecting on the accuracy of the information they receive. By performing a controlled experiment where they ask users to pay more attention to the accuracy of the information, it can reduce the spread of misinformation.

\noindent \textbf{Improve Media Literacy}
Media literacy refers to ``the ability to access, analyze, evaluate and create messages across a variety of contexts \cite{livingstone2004media}.'' This ability varies according to the medium of platforms, from reading and writing articles on traditional media to creating and sharing posts and videos on social media. Unlike media literacy for traditional social media where users engage in passive consumption, social media users actively create and share media content \cite{chen2011unpacking}. Therefore, the interactive environment of social media adds complexity to improving media literacy among social media users. Hung et al. \cite{hung2021new} discussed the vital role of media literacy in protecting people from health misinformation on social media. They found that media literacy had a positive relationship with COVID-19 preventive behaviors for adults in Taiwan. Similarly,  Roozenbeek et al. \cite{roozenbeek2022countering} found that improving media literacy among the public boosts people's ability to identify misinformation.

\noindent \textbf{Improve Health Literacy}
Apart from media literacy, health literacy adds more resilience to social media users against health misinformation. Health literacy refers to the ability to obtain, process, understand, and communicate health-related information needed to make informed health decisions \cite{berkman2010health}. Schulz and Nakamoto \cite{schulz2022perils} highlighted the importance of improving health literacy to counteract the effect of health misinformation. Specifically, they suggest the public broaden their health literacy beyond just reading and understanding medical information. Instead, health literacy should include more complex abilities, including understanding, accessing, applying, and appraising health knowledge.

% \cy{\cite{janmohamed2021interventions, janmohamed2022interventions}\cite{micallef2020role}}

%%%%%%%%%%%%%%%%%%%%%%%%%%%%%%%%%%%%%%%%%%%%%%%%%%%%%%%%%%%%%%%%%%%%%%%%%%%%%%%%%%%%%%%%%%%%%%%%%%%%%%%%%%%%%%%%%%%%%%%%%%%

\section{Datasets and Tools}
In this section, we will describe the relevant datasets related to healthcare misinformation, and introduce some existing tools that are developed to facilitate the combating effort of healthcare misinformation.
% In this section, we introduce the existing datasets and tools for online health misinformation detection. 

\subsection{Datasets}

\begin{table}[!htbp]

\caption{Recent Representative Datasets for Health Misinformation.}
\resizebox{\textwidth}{!}{%
\begin{tabular}{@{}llllll@{}}
\toprule
Dataset      & Health Topics                                                                                         & Instances & Sources                                                                                                      & Language                                                 & Year \\ \midrule
\texttt{SciTweets \cite{hafid2022scitweets}}    & general scientific                                                                             & 1,261     & Twitter                                                                                                     & English                                                  & 2022 \\ \midrule
\texttt{SCIFACT \cite{wadden-etal-2020-fact}}     & biomedicine                                                                                    & 50K       & PubMed                                                                                                      & English                                                  & 2020 \\ \midrule
\texttt{SCIFACT-OPEN \cite{wadden2022scifactopen}} & biomedicine                                                                                    & 500K      & S2ORC, arXiv, PubMed                                                                                        & English                                                  & 2022 \\ \midrule
\texttt{PUBHEALTH \cite{kotonya2020explainable}}   & \begin{tabular}[c]{@{}l@{}}biomedical subjects,\\ health-care policy\end{tabular}              & 11.8K     & PubMed                                                                                                      & English                                                  & 2020 \\ \midrule
\texttt{HEALTHVER \cite{Sarrouti2021EvidencebasedFO}}   & COVID-19                                                                                       & 1,855     & \begin{tabular}[c]{@{}l@{}}scientific evidence \\ from PubMed\end{tabular}                                  & English                                                  & 2020 \\ \midrule
\texttt{FakeHealth \cite{dai2020ginger}}  & \begin{tabular}[c]{@{}l@{}}16 health topics\\ (e.g. Alzheimer,)\end{tabular}                   & 2,296     & HealthNewsReview.org                                                                                        & English                                                  & 2020 \\ \midrule
\texttt{HealthLies \cite{Chaphekar2022HealthLiesDA}}  & \begin{tabular}[c]{@{}l@{}}Cancer, HIV/AIDS\\ Vaccine, SARS\\ ZIKA, Polio\\ Ebola\end{tabular} & 11K       & \begin{tabular}[c]{@{}l@{}}CNN, New York Times,\\ New Indian Express\end{tabular}                           & English                                                  & 2022 \\ \midrule
\texttt{RedHOT \cite{wadhwa2022redhot}}      & \begin{tabular}[c]{@{}l@{}}24 health topics \\ (e.g. ADHD, Diabetes)\end{tabular}              & 22K       & Reddit                                                                                                      & English                                                  & 2022 \\ \midrule
\texttt{MM-COVID \cite{li2020toward}}    & COVID-19                                                                                       & 23K       & \begin{tabular}[c]{@{}l@{}}Fact-checking websites\\ including Snopes, \\ Poynter, etc\end{tabular}          & \begin{tabular}[c]{@{}l@{}}Multi-\\ lingual\end{tabular} & 2020 \\ \midrule
\texttt{CoAID \cite{cui2020coaid}}       & COVID-19                                                                                       & 4,251     & \begin{tabular}[c]{@{}l@{}}Fact-checking websites\\ including LeaderStories,\\ PolitiFact, etc\end{tabular} & English                                                  & 2020 \\ \midrule
\texttt{CoVaxNet \cite{jiang2022covaxnet}}    & COVID-19 Vaccine                                                                               & -         & Online and Offline Data                                                                                     & English                                                  & 2022 \\ \midrule
\texttt{WICO \cite{schroeder2021wico}}        & COVID-19 Conspiracy                                                                            & 3,000     & Twitter                                                                                                     & English                                                  & 2021 \\ \midrule
\texttt{MMCoVAR \cite{chen2021mmcovar}}    & COVID-19 Vaccine                                                                               & 2,593     & \begin{tabular}[c]{@{}l@{}}News media include ABC,\\ CBC, The New York Times\end{tabular}                   & English                                                  & 2021 \\ \bottomrule
\end{tabular}
}
\label{data}
\end{table}

Although general misinformation datasets are widely available, there are only a number of datasets related to online health misinformation. Table \ref{data} shows a collection of publicly available datasets related to health topics such as diseases, biomedical subjects, vaccines, health conspiracy, and healthcare policy. Here are publicly available datasets related to health misinformation:
\begin{itemize}
    \item \textit{SciTweets \cite{hafid2022scitweets}:} This dataset contains annotated ground truth for science discourse on Twitter, such as COVID-19. Specifically, the dataset is divided into the following four different categories: scientific knowledge, reference to scientific knowledge, related to scientific research in general, and not science-related. SciTweets contains 1,261 human-annotated tweets and the consolidated ground-truth label for each category. 
    \item \textit{SCIFACT \cite{wadden-etal-2020-fact}:} This dataset contains 1,409 scientific claims paired with evidence-containing abstracts that are annotated with labels and rationales. The corpus is constructed on top of 5,183 abstracts from well-known journals like \textit{Cell, Nature, JAMA}.
    \item \textit{SCIFACT-OPEN \cite{wadden2022scifactopen}:} This is an expansion of the original SCIFACT dataset for open-domain scientific claim verification. The corpus contains 500K abstracts compared to the 5K abstracts from the original SCIFACT dataset.
    \item \textit{PUBHEALTH \cite{kotonya2020explainable}:} This is the first fact-checking dataset in the public health domain. Specifically, PUBHEALTH contains 11.8K claims along with gold-standard explanations by journalists related to health topics including biomedical subjects (e.g. infectious diseases, stem cell research), government healthcare policy (e.g. abortion, mental health), etc.
    \item \textit{HEALTHVER \cite{Sarrouti2021EvidencebasedFO}:} This dataset contains 14K evidence-claim pairs for fact-checking health-related claims. Specifically, the authors of HEALTHVER found that the proportion of complex claims in HEALTHVER is consistently higher than in the existing datasets.
    \item \textit{FakeHealth \cite{dai2020ginger}:} This is the first comprehensive fake health news dataset. It contains news content and news reviews from \textit{HealthNewsReview.org}, a fact-checking organization for health-related news from U.S. mainstream media. Specifically, it contains news content with rich features (e.g. text, image, tags), news reviews with detailed explanations (e.g. labels, explanations, news URL) for fake news detection tasks, social engagements, and a user-user social network.
    \item \textit{HealthLies \cite{Chaphekar2022HealthLiesDA}:} This dataset contains 11K facts and myths about diseases such as COVID-19, Cancer, Polio, Zika, HIV/AIDS, SARS, and Ebola.
    \item \textit{RedHOT \cite{wadhwa2022redhot}:} This dataset contains a corpus of 22K richly annotated social media posts from Reddit. It covers 24 health conditions including dysthymia, psychosis, thyroid cancer, ADHD, diabetes, etc. Each post is annotated with Populations, Interventions, and Outcomes.
    \item \textit{MM-COVID \cite{li2020toward}:} This dataset contains COVID-19-related news from 6 different languages including, English, Spanish, Portuguese, Hindi, French, and Italian. The news in MMCOVID dataset has news content, social engagements, and spatial-temporal information.
    \item \textit{CoAID \cite{cui2020coaid}:} This dataset contains 4,251 news along with their social engagements from fact-checking websites such as Politifact, and LeaderStories.
    \item \textit{CoVaxNet \cite{jiang2022covaxnet}:} This is a multi-source, multi-modal, and multi-feature online-offline dataset regarding COVID-19 vaccines. For online data, it contains social media data and fact-checking data. For offline data, it contains COVID-19 statistics, U.S. census bureau data, government responses, and local news.
    \item \textit{WICO \cite{schroeder2021wico}:} This dataset contains 3,000 manually annotated tweets related to the COVID-19 conspiracy theory that 5G technology is causally connected to the CVOID-19 pandemic.
    \item \textit{MMCoVAR \cite{chen2021mmcovar}:} This multimodal(images, text and temporal information) dataset contains 2,593 news articles from 80 publishers regarding the COVID-19 vaccine.
\end{itemize}

\subsection{Tools}
As discussed in Section \ref{intervention}, combating health misinformation requires collaborative efforts from other research disciplines. However, without proper UI, it is hard for researchers without machine learning backgrounds to apply detection models to their own research. Therefore, we want to highlight publicly available tools that can benefit researchers and different stakeholders.

\noindent \textbf{AVAXTAR} Schmitz et al. \cite{schmitz2021detecing} developed a Python package that analyzes Twitter profiles to assess their likelihood of sharing anti-vaccine information in the future. AVAXTAR system takes a Twitter ID and retrieves the target account's recent activity. It then calculates the probability of the target account's probability of posting future anti-vaccine information. Specifically, AVAXTAR uses three state-of-the-art models including Sent2Vec \cite{gupta2019better}, Sentence-MPNet \cite{song2020mpnet}, and Sentence-Distill-Roberta \cite{liu2019roberta}.

\noindent \textbf{QCRI's COVID-19 Disinformation Detector} To benefit both technical and non-technical end-users, Nakov et al. \cite{nakov2022qcris} developed a system to detect COVID-19-related misinformation. Specifically, the system contains publicly accessible APIs and a demo system that shows individual classified labels and aggregated statistics over time.

\noindent \textbf{COVID-19 Claim Radar} In order to track the vast variety of COVID-19-related claims, Li et al. \cite{Li2022COVID19CR} developed COVID-19 Claim Radar, a system that automatically extracts claims related to COVID-19 in news articles. COVID-19 Claim Radar provides users with a comprehensive structured view of COVID-19 claims, their associated knowledge elements, and related connections to other claims from news articles in multiple languages.

\section{Open Issues and Future Directions} \label{sec:open}
In this section, we present some open issues in combating health misinformation and future research directions. Health misinformation in social media is a newly emerging research area and an important public health issue, so
we aim to point out promising research directions that encourage collaborative and  multidisciplinary efforts.
\subsection{Interdisciplinary Research on Health Misinformation}
There is a pressing need to conduct interdisciplinary research to advance health misinformation characterization, detection, and intervention. For \textit{characterization}, it is important to understand why social media users are susceptible to health misinformation and why they spread it, which can be possibly explained by relevant social and psychological theories~\cite{uscinski2020people, van2022misinformation, tsirintani2021fake, zhou2021characterizing,scheufele2019science,dahlstrom2021narrative}.  Data-driven approaches are also important because of the capacity to derive statistical correlations between the circulation of health misinformation and user factors such as their demographic attributes (e.g., gender, race, age, ethnicity, religion, sexual orientation, etc), geographical regions, and temporal variations~\cite{luengo2021artificial}. Thus, joint efforts from social scientists, psychologists and data scientists are required to better characterize the causes and effects of health misinformation dissemination. Besides, previous research works mainly focus on the characterization of health misinformation circulation on conventional social media platforms such as Twitter and Facebook. A multi-channel analysis that includes more emerging platforms such as TikTok should attract more attention. For \textit{detection}, machine learning algorithms for misinformation detection have shown promising performances~\cite{kumari2021debunking, di2022health, aranacatania2022natural}. However, a large portion of current health misinformation datasets are only centered around several diseases (e.g., Covid, cancer, etc). Collecting misinformation datasets on more diseases and appropriately benchmarking the detection methods are strongly needed to assess the advancements of various model designs. We also observed that purely data-driven approaches can be limited by the size and quality of datasets and leveraging social theories to guide the model developments is  likely to further boost the detection performances~\cite{tang2014mining,shu2017fake}. For example, Shu et al.~\cite{shu2020detecting, dou2021user} have proposed several social theory-guided misinformation detection algorithms that pave the way toward this important direction. Moreover, considering the detrimental effects of health misinformation~\cite{west2021misinformation}, more research is desired on how to detect health misinformation in the early stage of dissemination.  For \textit{intervention}, it is important to evaluate the effectiveness of intervention measures in the real world~\cite{porter2021global}. Thus, it is important for public health analysts, machine learning experts, managers of social media platforms, and policy-makers to work together to investigate the impact of a variety of intervention actions and improve their practical functions among various groups of people~\cite{alam2020fighting,swire2019public,janmohamed2021interventions, janmohamed2022interventions}. 
% it is important to understand the causes of misinformation dissemination. Public health analysts and machine learning experts can work together to better investigate and evaluate the causal factors of whether specific types of users are more likely to spread health misinformation, and assess the causal impacts of online health misinformation and offline public health crisis. 

% \cy{dataset \& benchmark}

% \cy{Early Health Misinformation Detection and Intervention}
% \cite{10.1145/3511808.3557263}

% \cy{from psychological perspective, why social media users are susceptible to health misinformation\\
% from sociological perspective, the role of government, social media platforms, information consumers, experts, influencers, malicious uers in the dissemination of health misinformation\\
% from statistical perspective, the statistical correlation between the circulation of health misinformation and factors such as demographic attributes （e.g., gender, race, age, ethnicity, religion, sexual orientation, etc）, geographical regions and temporal variations \\
% from data-driven perspective, utilize artificial intelligence and machine learning methods to trace the dynamics of health misinformation on social media\\
% from causal inference perspective, causal effect analysis of Health Misinformation,
% the causal relation between Health Misinformation \& real-world events}

% \subsection{Trustworthy/Responsible/human-in-the-loop Misinformation Detection}
\subsection{Trustworthiness in Combating Health Misinformation}

Although artificial intelligence-based health misinformation detectors have achieved relatively high performances~\cite{paraschiv2021unified, zhao2021detecting, tashtoush2022deep,whrl2022entitybased}, the trustworthiness in combating health misinformation has attracted much less attention. However, to make sure that artificial intelligence models do not cause unintentional harm to humans and gain the public's trust, it is more and more important to take trustworthy properties into consideration including safety, robustness, fairness, explainability, privacy, and accountability~\cite{liu2021trustworthy}. Some initial efforts mainly focus on the explainability aspect~\cite{ayoub2021combat,10.1145/3492855,10.1145/3485447.3512257,10.1145/3394486.3403092,kotonya2020explainable,kou2022crowd}. For example, Ayoub et al. ~\cite{ayoub2021combat} proposed a DistilBERT~\cite{sanh2019distilbert} based detection model and adopted SHapley Additive exPlanations (SHAP)~\cite{DBLP:conf/nips/LundbergL17} to improve model explainability. Some works bring biomedical knowledge to enhance the interpretability~\cite{10.1145/3492855,10.1145/3394486.3403092,kotonya2020explainable}.  Kou et al.~\cite{kou2022crowd} proposed a crowd-expert-AI framework to jointly leverage crowd workers and health experts to generate natural language explanations for the Covid misinformation detector.  In addition, Shang et al.~\cite{10.1145/3485447.3512257} leverage multi-modal social context to generate explanations for model predictions. Although these works already conduct some studies on the explainability property of health misinformation detection, it is still under exploration on combing misinformation contents, user engagements, social structures, and external knowledge to generate convincing model explanations. Besides, it is also essential to study fairness, safety, robustness, privacy, and accountability in health misinformation detection, which needs further research in the future. Lastly, there start to be some research on the intersection of different trustworthy properties in machine learning~\cite{chen2022fair,xu2020robust,lamy2019noise,liang2022joint}, it is worth studying on building health misinformation detectors that satisfy multiple trustworthy properties simultaneously. 

% fairness~\cite{park2021presence} \cite{sahadevan2022gender}

% \cite{chang2022disparate}

% explainable
% \cite{ayoub2021combat}
% \cite{10.1145/3492855} \cite{10.1145/3485447.3512257}

\subsection{Human-centred Health Misinformation Analysis}

Human-centered health misinformation analysis refers to \textit{bringing humans in the loop} in online health misinformation analysis, the goal of which is to make sure the process of characterization, detection, and intervention of health misinformation is beneficial to humans and would not cause unintended harm to specific groups. Generally, there are four core dimensions of human-centered health misinformation analysis: why humans are susceptible to health misinformation, how humans spread health misinformation, how to interact with humans for detecting health misinformation, what intervention measures against health misinformation are effective and beneficial for humans~\cite{van2022misinformation}. For the \textit{susceptibility} aspect,  some initial efforts analyze potential variables influencing humans' susceptibility to health misinformation in social media. For example, Pan et al.~\cite{pan2021examination} discovered that some demographic factors (e.g., gender, age, income) largely contribute to the acceptance of online health misinformation. Besides, there are some research works tackling the problem of humans' susceptibility from a psychological perspective~\cite{uscinski2020people, scherer2021susceptible,pennycook2021psychology}. However, the existing research works cannot fully characterize the whole picture of humans' susceptibility to health misinformation and more analysis from perspectives such as sociology, politics, culture, beliefs, and religions is desired.
For the \textit{sprend} aspect, some researchers leverage the susceptible–Infected–recovered (SIR) model from epidemiology to model the spread of health misinformation in social media~\cite{cinelli2020covid,juul2021comparing}. However, this line of work does not fully take human factors into consideration because people exposed to health misinformation unnecessarily would believe it. More research  is needed to obtain a better understanding of the association between exposure to health  misinformation  and  persuasion~\cite{van2022misinformation}.
For the \textit{detection} aspect, some works have conducted initial explorations on \textit{keeping humans in the loop} to make health misinformation detectors more interpretable~\cite{kou2022crowd}. However, it is still under study on how to leverage interaction with humans to achieve more trustworthy properties including fairness, robustness, privacy, safety, and accountability in the algorithm design of health misinformation detection. On the other hand, it is urgently desired to evaluate the effectiveness and impacts of various health misinformation detectors from a human perspective.
For the \textit{intervention} aspect, the standard measure for preventing health misinformation is fact-checking and debunking after humans are exposed to it~\cite{paynter2019evaluation,yousuf2021media,paynter2019evaluation,walter2018unring,chan2017debunking,walter2021evaluating}. However, these post-hoc intervention methods potentially would cause ``backfiring effect'' that makes humans end up believing more in the health misinformation~\cite{lewandowsky2012misinformation,swire2020searching}. Thus, a joint effort from psychologists, social media platforms, and policy-makers is needed to design more psychological inoculation methods and policies and boost humans' immunity against online health misinformation.

% human psychology

% take human in the loop

% Build human's trust

% \cite{vaughan2020human} A Human-Centered Agenda for Intelligible Machine Learning

% human-in-the-loop
% \cite{kou2022crowd}

% \cy{human \& misinformation,
% human \& health}

% \cy{tackle the problem from human perspective}

% \subsection{Online and Offline Health Misinformation}
\subsection{Online and Offline Correlation for Health Misinformation}

Although health misinformation is widely spread in various online platforms, it has been demonstrated to be deeply associated with the offline physical world~\cite{zhang2022counterfactual,world2021infodemic}. However, existing works pay more attention to understanding online misinformation and offline events separately~\cite{darwish2022survey}. An integrated analysis is strongly desired to investigate the real connections between online and offline space.
Generally, the online and offline correlation for health misinformation embraces two perspectives: \textit{how the online dissemination of health misinformation impacts offline real-world events} and \textit{how offline actions can help mitigate the harm of online health misinformation}. For the \textit{first} aspect, there are already some empirical studies on the real-world impact of health misinformation. For example, increasing works studied the vaccine hesitancy phenomenon impacted by online vaccine misinformation~\cite{getman2018vaccine,wilson2020social,samal2021impact,kanozia2021fake}. Chary et al.~\cite{chary2021geospatial} discovered a strong statistical association between COVID-19 health misinformation and poisoned patients with household cleaners. More broadly, online health misinformation may deepen the ``inequality-driven mistrust'' among minority groups and lead them to deter or refuse seeking medical care in the Covid pandemic, which would cause increasing health risks and disproportionate harm to the already-vulnerable populations~\cite{luengo2021artificial}. What's worse, the mistrust against the government will make these minority groups more likely to accept online health rumors, which forms a vicious circle of health misinformation among these groups~\cite{jaiswal2020disinformation}. However, more dimensions for the real-world effects of online health misinformation are under exploration: (1) how to conduct evidental experiments in the physical world to verify the impacts of online health misinformation; (2) how to investigate the causal correlation between health misinformation and real-world events; (3) how to analyze the offline effects of health misinformation from a sociological perspective.  For the \textit{second} aspect, social media platforms and policy-makers have made significant efforts on designing a variety of measures to prevent the dissemination of online health misinformation~\cite{shattock2021self,arnold2021source,MVincent2022,dias2020emphasizing,pomeranz2021governmental}. 
However, there are still some important problems to be solved. For example, more appropriate analysis on the effectiveness of these approaches on actual dissemination of online health misinformation is needed. More importantly, the evaluation of whether or not these actions can benefit minorities and underrepresented groups of people and break the vicious circle of online health misinformation among them needs more effort in the future.

% The spread of health misinformation on social media has serious offline consequences that put people's health in danger. Chary et al. \cite{chary2021geospatial} showed evidence that links online advice of using bleach as COVID-19 prophylaxis and a spike in bleach poisoning. Therefore, we need to provide both online detection and offline intervention solutions. For online detection, it is essential to identify the causal effect of health misinformation. Some preliminary work on this includes \cite{zhang2022counterfactual}, where Yizhou et al. developed a causal framework that model the causal influence of misinformation on social media using temporal point process. However, this area still needs to be explored extensively. The lack of high quality data also poses great challenge to research in this area. For offline intervention, improving public media and medical literacy is important. However, this requires a joint effort from health officials, media platforms, and public leaders.

% \cy{how online impact on offline? how offline impact on online?}

% \cite{shu2022combating}

% \cite{verma2022examining}

% \cite{zhang2022counterfactual}

% how online impact on offline?

% \cite{sylvia2020we}

% vaccine hesitancy

% \cite{getman2018vaccine}
% \cite{wilson2020social}

% how offline impact on online?

% \cite{helmi2018community}

% \cite{armfield2007public}
% \cite{arcus1977misinformation}

\subsection{Health Misinformation and Health Disparity}

Health misinformation and health disparities are both prominent concepts in public health literature but the correlation between them is under study. 
Health disparities refer to the differences in health status across groups that defined by factors such as gender, race, ethnicity, disability, income, sexual orientation, or geographic locations~\cite{bartley2016health}.   Previous studies have shown that people of color and underserved groups face longstanding health disparities~\cite{bailey2017structural,braveman2011social,gee2011structural,braveman2006health,martin2019clinical,zavala2021cancer} and the COVID-19 pandemic further worsened the existing inequities~\cite{artiga2020disparities,chowkwanyun2020racial,lopez2021racial,nana2021health,greenaway2020covid}.
% \ks{These papers are published long time ago, and how come they are talking about covid}. 
However, there is very limited research on the association between the proliferation of health misinformation on social media and health disparities~\cite{southwell2022health}. Some research works point out that the demographic factors contribute to the acceptance of online health misinformation for different groups of people~\cite{pan2021examination,luengo2021artificial,paakkari2020covid,jaiswal2020disinformation}. For example, Pan et al.~\cite{pan2021examination} found out that individuals with lower socioeconomic status are more likely to accept health misinformation compared to those from higher socioeconomic groups. Thus, the proliferation of health misinformation in social media would cause more harm to people from low socioeconomic groups and deepen the existing health disparities. More seriously, since a large portion of health misinformation involves fake cures, invincibility rumors, or ineffective preventive medical measures, those already-vulnerable populations are more likely to be led to  health-related risky behaviours~\cite{luengo2021artificial}. Thus, it is increasingly critical to study the role online health misinformation plays in explaining the documented health disparities and investigate more intervention measures to mitigate the effects of exposing to health misinformation among groups that face relatively worse health outcomes~\cite{southwell2022health}. 
Moreover, since artificial intelligence has been widely used in various applications on health misinformation characterization, detection, and intervention~\cite{wadden2021multivers,zhang2022counterfactual}, the intrinsic bias of AI itself may cause disproportionate dissemination of health misinformation among minority groups and runs the risk of exacerbating the health inequalities~\cite{luengo2021artificial}. Thus, more efforts are needed to overcome the negative impacts of AI when utilizing it to combat health misinformation in social media.

\section{Conclusion}
% In this paper, we highlight the importance and pressing need for studying health misinformation in social media, and proposed open issues and challenges for future research in this field. First and foremost, we want to emphasize that combating health misinformation in social media is a multidisciplinary issue. We hope this work can provide a catalyst to inspire multi-discipline work toward research on this topic. In conclusion, we first systematically reviewed existing research literature from three perspectives: characterization, detection, and intervention. For characterization, we examined the cause, effect, influence strategy, and dissemination patterns of online health misinformation. For detection, we categorized the existing detection models into six classes including capturing health content, exploiting user engagements, modeling social structures, incorporating biomedical knowledge, connecting multiple modalities, and leveraging other domains. For intervention, we discuss the intervention efforts needed from the government, organizations, and individuals. Finally, we identified the pressing open issues and future research directions on health misinformation in social media.

The increasing popularity of social media has enabled the proliferation of online health misinformation, which has brought dramatic impacts on the public and society. In this survey, we conduct a comprehensive review of existing research on health misinformation in social media from different disciplines. We also systematically organized the related literature in three phases: characterization, detection and intervention. For characterization, we examined the causes, effects, influence strategies, and dissemination patterns of online health misinformation. For detection, we categorized the existing detection models into six classes including capturing health content, exploiting user engagements, modeling social structures, incorporating biomedical knowledge, connecting multiple modalities, and leveraging other domains. For intervention, we reviewed the efforts from different levels including the government, organizations, and individuals. In addition, we provided existing datasets and tools on analyzing, detecting and preventing online health misinformation.  Finally, we discussed the pressing open issues and future research directions on combating health misinformation in social media.

% \section{Related surveys}
% \ks{we remove this section and rephrase the content and add to intro}
% \cite{darwish2022survey}
% \cite{di2022assessing}
% \cite{vyas2021proliferation}
% \cite{cabitza2022responsible}
% ~\cite{tsirintani2021fake}
% survey on Misinformation and public opinion of science and health\cite{cacciatore2021misinformation} (PNAS)

% \cite{van2022misinformation} [nature medicine]

% \cite{gupta2020information}

% \cite{suarez2021prevalence}

% \cite{ding2020challenges}

% \cite{wang2019systematic}

% ~\cite{sanaullah2022applications}

% \cy{\cite{alam2020fighting}}

% \cy{\textbf{references from PNAS} \cite{west2021misinformation}
% \cite{scheufele2019science}
% \cite{dahlstrom2021narrative}
% \cite{porter2021global}}

%%%%%%%%%%%%%%%%%%%%%%%%%%%%%%%%%%%%%%%%%%%%%%%%%%%%%%%%%%%%%%%%%%%%%%%%%%%%%%%%%%%%%%%%%%%%%%%%%%%%%%%%%%%%%%%%%%%%%%%%%%%

\bibliographystyle{ACM-Reference-Format}
\bibliography{main}

\end{document}